\documentstyle[prl,aps,epsfig]{revtex}

\begin{document} 
\twocolumn[
\hsize\textwidth\columnwidth\hsize\csname@twocolumnfalse\endcsname

\title{Critical exponents in spin glasses :  numerics and experiments}
\author{I.A. Campbell, D. Petit, P.O.Mari}

\address{Laboratoire de Physique des Solides,\\ Universit\'e Paris Sud, 91405
Orsay, France}
\author{L.W. Bernardi}
\address{Institute for Solid State Physics , University of Tokyo,\\
7-22-1 Roppongi, Minato-ku, Tokyo 106-8666}

\maketitle 

\begin{abstract}

We give an overview of numerical and experimental estimates of critical
exponents in Spin Glasses.  We find that the evidence for a breakdown of
universality of exponents in these systems is very strong.
\end{abstract}

\twocolumn
\vskip.5pc ] 
\narrowtext




\sloppy


\section{Introduction}

It is well known that at a continuous phase transition, striking critical
behaviour is observed.  As the transition temperature is approached from above
or below, there are power law singularities in a number of physical parameters (
the specific heat, the susceptibility,...).  On very general grounds it can be
shown that various critical exponents which govern the singularities are related
to each other through scaling relationships.  Even more remarkable is the fact
that systems which are very different from each other at the microscopic level
can be arranged into universality classes :  within a given class all members
have strictly identical exponents.  Classes are defined by a restricted number
of parameters - basically the space dimension and the number of components of
the order parameter of each system.  For standard second order transitions this
behaviour can be fully understood, and the exponents calculated {\it a priori}
thanks to renormalization group theory.  This is one of the outstanding
achievements of statistical physics (see Fisher \cite{fisher} for an
enlightening historical survey).

At standard ferromagnetic transitions, the exponents follow mean field behaviour
down to dimension 4 (the upper critical dimension).  At dimensions $d$ lower
than 4, the exponents have been calculated though the $\epsilon$ expansion, 
where
$\epsilon = 4-d$.  There are well behaved series in increasing powers of 
$\epsilon$
which allow one to give renormalization group estimates of the exponents.  The
values can be compared with those measured from high precision numerical
simulations, and contact can be made with the exact values at dimension 2 (where
for the Ising system, the exponents have been known exactly for more than $50$
years\cite{ons}).  Renormalization group values, numerical values, and
experimental values are all in excellent agreement with each other.  The only
exponent whose value is not quite so thoroughly established is the dynamic
exponent $z$ ; even here disagreements between different estimates are small.
Universality is a general rule in systems with standard second order
transitions, except for a restricted class of two dimensional systems with
conflicting interactions.  Here it has been shown analytically as well as
numerically that the exponents vary continuously as a function of the ratio of
the competing interactions \cite{bax,suz}.  This behaviour can be explained for
these particular systems in terms of marginal operators in the renormalization
scheme \cite{kad}.

In the Spin Glass (SG) context, for a long time it was by no means obvious that
there were well defined phase transitions at all in real three dimensional
materials, or even in 3d model systems.  Although the cusp temperature is
clearly marked experimentally, the specific heat shows no visible singularity,
and the susceptibility does not diverge in the region of the cusp temperature.
A very important step forward was the realization in the in the late 1970s 
that the
appropriate parameter to measure is not the standard linear susceptibility but
the non-linear susceptibility \cite{suz2}.  The Miyako group in Sapporo in
pioneering work \cite{miyako} showed that there is a divergence of $\chi_{nl}$
at the cusp temperature ; this work and that of other groups
\cite{monod,barbara,omari,beauvillain,vincent} convinced the community that
there is indeed a {\it bona fide} transition in a SG.  With the existence of a
transition established, estimates were found of the critical exponents.
Numerical work soon followed \cite{singh,BY,og}.  It was important to measure
these values in order to compare with the renormalization group approach.  Does
the combination of frustration and randomness which characterize SGs modify the
basic physics of transitions in a fundamental way or not ?  This has turned out
to be a long story, which is still not finished.  Unfortunately, as no clear
theoretical guidelines appeared, the enthusiasm for the subject dropped and even
the empirical ground rules were not fully established.  Our aim here is to show
that this study, both numerical and experimental, is well worth pursuing.

Before looking in detail at the data, we can first note that the exponents which
come out of the numerical or experimental analyses are very different from the
classical ferromagnetic values.  For the latter, in dimension 3 (for either
Ising, XY, or Heisenberg spins) the exponents $\alpha$ and $\eta$ are
numerically small, and $z$ is near $2$.  For the spin glasses $\alpha$ is
strongly negative (around $-2$ which is consistent with the lack of a visible
singularity in the specific heat )and $\eta$ is far from zero ; experimental
values range from $+0.4$ to $-0.5$ and simulation values are always distinctly
negative.  $z$ is strong - generally around $6$.  The major differences compared
with the standard second order transition values already indicate that the spin
glass transition lies in a quite different category.

In Ising Spin Glasses (ISGs) the upper critical dimension is $6$ and one could
imagine that from dimension $6$ down, a similar renormalization group approach
would be valid {\it mutatis mutandis} as for the ferromagnets.  In fact things
are much more complicated.  The $\epsilon$ expansion has been calculated to 
order
three \cite{green}, but the successive factors in increasing powers of epsilon
grow rapidly so it is not at all obvious how the total series will sum.  Valiant
efforts have been made over many years to set the theory on a firm footing using
field theory \cite{cyrano}, but so far the only clear result is to confirm that
the leading term in $6-d$ from the $\epsilon$ expansion is correct. 
 After that, numerous badly controled
terms proliferate and theory is of little practical help in predicting exponents
even at $d = 5$.

As a predictive theory is lacking, we are forced to turn to numerical and
experimental methods so as to establish the empirical values of the exponents.
The empirical results show clear violation of the Universality rules.

\section{Numerical results}

In the following discussion we will concentrate mainly on the systems which have
been studied the most fully - ISGs on (hyper)cubic lattices with random unbiased
interactions between near neighbours.

The definitions of the critical exponents are familiar, with appropriate
modifications for spin glasses to take into account the fact that that the order
parameter is the Edwards Anderson parameter.  The specific heat exponent is
$\alpha$, the order parameter exponent below the ordering temperature is
$\beta$, the spin glass susceptibility exponent is $\gamma$.  The non-linear
magnetization at the ordering temperature exponent is $\delta$, with
\begin{equation} M_{nl} \sim H^{2/\delta} \end{equation}.  The correlation 
length
exponent is $\nu$, and the exponent for the form of the correlation function at
the ordering temperature is $\eta$.  The relaxation time dynamic exponent is
$z$.  The scaling relationships between these exponents are 
$\alpha+2\beta+\gamma=2$, $d\nu = 2 - \alpha$, $\gamma = (2-\eta)\nu $,
 $\nu(d-2+\eta)=2\beta$ and 
$\delta=(d+2-\eta)/(d-2+\eta)$. Numerically each of the exponents
can in principle be measured independently through temperature dependences on
large samples , though frequently they are measured through using finite size
scaling relationships.  Experimentally, $\gamma$,$ \alpha$, $\delta$ and the
combination $z\nu$ can be measured directly while $\beta$,$\nu$ and $\eta$ can
only be obtained through scaling.  There are many other useful relationships ;
for instance the relaxation of the autocorrelation function at $T_g$ has an
exponent \begin{equation} x=(d-2+\eta)/2z \end{equation}

In addition to the standard critical exponents other exponents can be defined,
in particular the stiffness exponent $\theta$.  This is not a critical exponent
but is defined by the size dependence of changes in energy with boundary
conditions\cite{BM}.  For a spin glass energy measurements can be made with
periodic and anti-periodic boundary conditions.  The variation of the sample to
sample fluctuations of the energy differences scale as $L^\theta$.  If the zero
temperature $\theta$ is positive, then the ordering temperature is greater than
zero.

A first type of quasi-numerical approach is furnished by high temperature
expansions\cite{singh,singh2,klein}.  The method consists of an extrapolation
from a finite number of exact terms in the high temperature series expansion of
some thermodynamic function to its asymptotic coefficients.  The asymptotic form
of the series contains the information on the singularities of the function.
The extrapolation is not exact, but excellent results have been obtained in
regular systems.  The situation is less favourable in disordered systems.
Careful analysis of the spin glass susceptibility from a series with a large
number of terms (up to 20 in dimension 3) provides a set of estimates for the
values of $T_g$ and $\gamma$ obtained from different approximant functions.  If
the series expansion was infinite, the method would become exact but in practice
the limited length of the series means that the estimates are not perfect.  The
exponent results become less accurate as one gets further from the upper
critical dimension.  For dimensions $5$ and $4$ the longest series give high
quality estimates which can be used as independent yardsticks to compare with
the Monte Carlo data which we will discuss below.  The method does not have the
same problems (such as thermalization) which are encountered in Monte Carlo
simulations, but considerable know-how is needed to calculate long series and to
extract reliable exponent estimates from the raw series results.  Up to now the
only specific ISG series expansion results published are for binomial ($\pm J$)
interactions.

For dimension $5$ a reliable and accurate ordering temperature and set of
exponents is given by \cite{klein}.  For dimensions $4$ and $3$ the results
quoted in \cite{singh2} are more transparent ; it is clear that the approximant
data points cluster satisfactorally with a strong correlation between estimates
for $T_g$ and those for $\gamma$.

The most widely used technique for determining critical exponents numerically
has been that of Monte Carlo simulation.  Many efforts have been made to measure
ISG critical exponents accurately, despite considerable technical difficulties.
For measurements exploiting the finite size scaling method \cite{BY}, each
sample is first annealed numerically until the spin system can be judged to be
in thermal equilibrium ; then the fluctuations in the autocorrelation function
$q(t)=<S_i(0)S_i(t)>$ are measured.  The precautions necessary are described in
\cite{BY}.  Long enough times must be used for each part of the procedure, and
the time scale defining " long enough " depends on the size $L$ and the
temperature $T$.  The larger the sample and the lower the temperature the longer
the time scales, so it becomes very difficult to obtain significant data on
large samples at low temperatures.  Sophisticated update methods have been
developed which aleviate this problem to some extent\cite{anneal}.  One must be
sure that measurements have been done over a sufficient number of independent
samples.  Even if numerically high quality data has been obtained, there may be
intrinsic corrections to finite size scaling which mean that the scaling rules
(which are valid asymptoticly for large sizes) may not yet hold exactly for the
range of sizes studied.

An important parameter which can be deduced from the equilibrium fluctuations is
the spin glass susceptibility $\chi_{SG} = <q^2>$, directly related to the
non-linear susceptibility.  The finite size scaling form is \begin{equation}
\chi_{SG} = L^{2-\eta} f(L^{1/\nu}(T-T_g)) \end {equation} Precisely at $T_g$,
as a function of size \begin{equation} \chi_{SG}(L) \sim L^{2-\eta}
\end{equation} $T_g$ can be determined quite accurately as the highest
temperature at which $log(\chi_{SG}/L^2)$ varies linearly with $log(L)$ up to
large $L$.

These expressions ignore corrections to finite size scaling.  There should be a
further factor so that for instance the spin glass suceptibility is multiplied 
by
 $[1-L^{-w}
f_L(L^{1/\nu}(T-T_g))+...]$. The correction to scaling exponent
$w$ has been estimated at around $2.8$ in the binomial ISG in dimension
3\cite{mari}, as against $0.9$ for the 3d Ising ferromagnet and $1.6$ for 3d
site percolation \cite{bal}.  It turns out that the strength of the correction
to scaling can change dramatically from one system to another.

Many authors estimate the critical temperature $T_g$ through the Binder the
cumulant method \cite{BY}.  The cumulant for a given $T$ and $L$ is defined by a
dimensionless combination of moments of the autocorrelation fluctuations
averaged over a large number of samples :  \begin{equation} g_L = 1/2(3 - <q^4>
/<q^2>^2) \end{equation} for Ising spins.  This cumulant is defined so as to go
from zero for a high temperature Gaussian distribution of $q(t)$ values, to $1$
for a unique low temperature state.  (Other related cumulants can be defined).
For a continuous transition with a critical temperature $T_g$, $g_L(T_g)$ should
be independent of size $L$, with values fanning out as a function of $L$ above
and below $T_g$.  Once $T_g$ has been established accurately, the exponents can
be estimated by plotting the whole set of data for $g_L(T)$ and for
$\chi{SG}(L,T)$ in an appropriate scaling form.

In practice, sample to sample fluctuations in $<q^4>$ are strong so very large
numbers of samples must be measured (the lack of self averaging is very much
worse for $<q^4>$ than for $<q^2>$).  The crossing point can be ill determined
because the $g_L(T)$ curves do not fan out appreciably at temperatures lower
than the crossing point.  When this is the case, relatively minor corrections to
finite size scaling can modify the apparent crossing temperature.  (Correction
factors of the form given above should apply to both $<q^4>$ and $<q^2>$).
Finally, the values of the exponents deduced from the scaling plots, strongly
correlated with the $T_g$ estimate, frequently vary very steeply with the
apparent value of $T_g$.  In conclusion, the exponent estimates obtained in this
traditional way must be treated with considerable caution, even when large scale
numerical efforts have been made.  Accurate and reliable results can be obtained
only in favourable circumstances.

When calculations have been made to large sizes, it is possible to use the
scaling rules for the spin glass susceptibility and for the spin correlation
length to estimate the critical temperature and the exponents by extrapolations
to infinite size \cite{matteo}.  This method should be reliable, as corrections
to finite size scaling are kept well under control.

Alternative techniques which have been less widely used rely (at least partly)
on dynamic measurements.  In massive simulations on large samples ($64^3$ spins)
Ogielski \cite{og} studied the relaxation of $q(t)$ as a function of 
temperature.
An advantage of this method is that a strict thermal equilibrium state is not
necessary ; as long as the anneal has been made over a time $\tau_a$ much longer
than the subsequent measuring time over which $q(t)$ is studied, the measured
$q(t)$ curve will be the true thermal equilibrium form (see for instance
\cite{rieger}).  Ogielski assumed the standard critical behaviour for $q(t)$,
which is \begin{equation} q(t) = t^{-x} f(t/\tau(T)) \end{equation} He estimated
$T_g$ from the divergence of the relaxation time $\tau(T)$.  With $T_g$ in hand
he estimated the critical exponents from a combination of dynamic and
equilibrium measurements.

One can also exploit the critical behaviour of strictly non-equilibrium
dynamics.  Suppose a spin glass sample is initially at infinite temperature (so
the spins have random orientations) ; it is then annealed from this
configuration to the critical temperature $T_g$.  The non-equilibrium spin glass
susceptibility at a time $t$ after the start of the anneal will increase as
$t^h$ \cite{huse}where \begin{equation} h= (2-\eta)/z \end{equation} Analagous
non-equilibrium dynamic parameters have been studied extensively in a large
number of regular systems \cite{zheng}.  This non-equilibrium scaling behaviour
has been established on a very firm theoretical basis \cite{jannsen}.  An 
obvious
practical advantage for numerical work is that no preliminary anneal is
required.

Now it is clearly possible to combine the measurements of the dynamic relaxation
exponent $x(T)$ and $h(T)$ at a series of test temperatures $T_i$ to obtain a
sequence of apparent exponents $\eta(T_i)$ and $z(T_i)$ :  \begin{equation} \eta
= (4x + (2-d)h)/(2x+h) \end{equation} \begin{equation} z = d/(2x + h)
\end{equation} The $\eta(T_i)$ can be compared with independent $\eta(T_i)$
estimates from the equilibrium spin glass susceptibility (equation (9)).
Consistency dictates that at the true $T_g$ the two estimates must coincide.
This leads to estimates of $T_g$, $\eta$ and $z$ which turn out in practice to
be precise (the crossing is clean ) and virtually free from pollution by
corrections to finite size scaling\cite{lwb}.  Once these parameters are well
established, the other exponents such as $\nu$ can be determined from
conventional scaling plots with only one unknown parameter.  We will refer to
this method as the " three scaling rule " technique.

The low dimensions present special cases.  First, in dimension $1$ where
critical temperatures are certainly always zero for short range interactions,
the stiffness exponent $\theta$ is exactly $-1$ for continous distributions of
interactions and $0$ for $\pm J$ interactions \cite{BM}.  In dimension $2$ it is
also well established that the ordering temperature $T_g = 0$ \cite{BY} (except
possibly when the interactions are $\pm J$ \cite{shira}).  From the definition
of $\eta$, when $T_g$ is zero and for a unique ground state (corresponding to
any continuous distribution of interactions such as the Gaussian distribution)
there is an additional scaling rule \cite{BY} \begin{equation} \eta = 2 - d
\end{equation} As fewer spins are involved for a given size L than at high
dimension, it is easier to cover a wide range of L for finite size scaling in
Monte Carlo simulations.  Even better, sophisticated numerical techniques exist
to find exact ground states for systems up to large sizes\cite{matteo2}.
(Recently exact ground states in dimension 3 have also been obtained to quite
large $L$\cite{alex}).

\section{Exponent values}

We will concentrate on the values of $\eta$, and start from the low dimensions.

For the 2d binomial ISG, $\eta$ has been estimated to be $0.20\pm
0.02$\cite{BY}.  Curiously, even if the $T_g$ is small but non-zero, the
estimated value of $\eta$ is very similar\cite{shira}.

\begin{figure}
\begin{center}
\epsfig{file=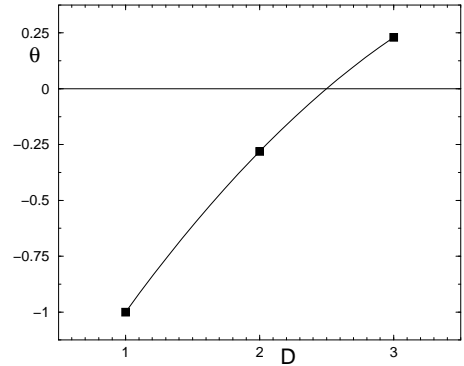,height=5cm}
\end{center}
\caption{Stiffness exponent $\theta$ as function of dimension for ISGs 
with Gaussain interactions. Dimension 1 \protect\cite{BM}, dimension 2 \protect\cite{rieger2},
 dimension 3 \protect\cite{matteo2}    .}
\label{fig:1}
\end{figure}

For the Gaussian distribution, from the values of the stiffness exponent
$\theta(d)$ as a function of dimension\cite{BM,rieger2,matteo2}, we can
estimate the lower critical dimension $d_l$ , the dimension at which $\theta=0$,
figure~\ref{fig:1}.  Because the Gaussian distribution is continuous, as $d_l$ is close to
$2.50$, $\eta(d_l)$ must be close to $-0.50$, from equation (10).

Numerous estimates have been given of the exponents for ISGs in dimension 3 with
binomial or Gaussian interaction distributions.  We have summarized the
situation in \cite{mari}, where we find that there are strong deviations from
finite size scaling for the binomial case and where we obtain reliable exponent
estimates principally from the three scaling rule method.  Recent data on the
binomial case established by extrapolation to infinite size \cite{matteo} are
consistent with the values of \cite{mari}.  Other estimates relying on the
Binder cumulant method are probably affected by corrections to finite size
scaling and so are less reliable.

We will use the results obtained in dimension 4 to demonstrate the coherence of
the different methods when these are used carefully, and the inescapable
conclusion that the values of the exponents change with the form of the
interaction distribution.

\begin{figure}
\begin{center}
\epsfig{file=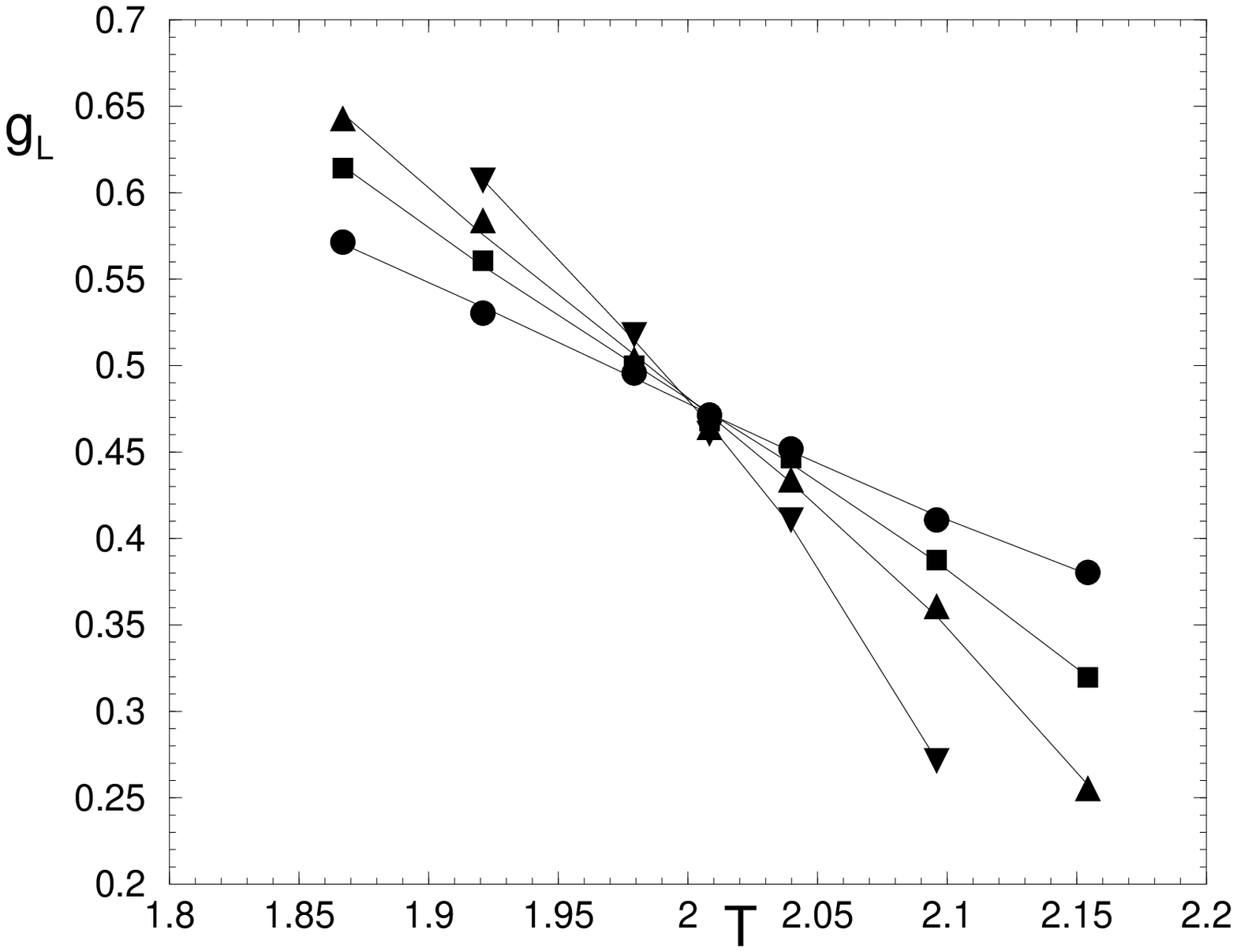,height=5cm}
\end{center}
\caption{Binder cumulant data for the binomial ($\pm J$) interaction ISG 
in dimension 4, (A.P. Young, unpublished). Sizes $L=4,6,8,12$. The curves should intersect at 
the critical temperature.}
\label{fig:2a}
\end{figure}

  Figure~\ref{fig:2a} shows high precision data for the Binder
cumulant in the binomial case\cite{apy}.  There is a clean intersection of the
curves at $T=2.00 \pm 0.01$.  

\begin{figure}
\begin{center}
\epsfig{file=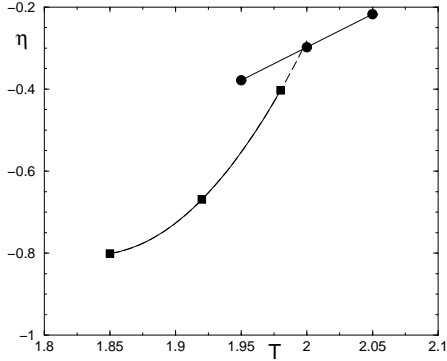,height=5cm}
\end{center}
\caption{$\eta(T)$ data for the binomial interaction ISG in dimension 4. 
$\eta_1$ (squares) is from the spin glass susceptibility (A.P. Young, unpublished) and
 $\eta_2$ (circles) is from the $h$ and $x$ effective exponent, see text. The 
 intersection should correspond to the critical temperature $T_g$  and $\eta(T_g)$.}
\label{fig:2b}
\end{figure}

Figure~\ref{fig:2b} shows the intersection of the two curves
for $\eta(T_i)$ used in the three scaling method described above\cite{lwb2}.
The intersection is at precisely the same temperature, validating this
technique.  From the intersetion point we can deduce $T_g$ and $\eta =0.30 \pm
0.02$.  As the slope of the curve for $\eta(T_i)$ for the estimate from the
dynamic exponents $x$ and $h$ is much weaker than from that deduced from the 
equilibrium finite
size scaling spin glass suceptibility, the precision on the value of $\eta$ is
much higher using this method rather than working only with the pure equilibrium
susceptibility and Binder cumulant data.  The series expansion data
\cite{singh2} is in quite independent agreement with these Monte Carlo results.
>From the results plotted in Figure 3 of \cite{singh2}, for this value of $T_g$
one would expect $\gamma = 2.1 \pm 0.1$, or $\nu = 0.92 \pm 0.05$ from the
scaling relation.  This value of $\nu$ is perfectly consistent with direct
scaling of the Monte Carlo spin glass susceptibility data.

\begin{figure}
\begin{center}
\epsfig{file=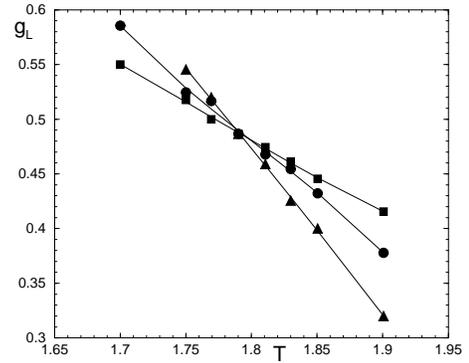,height=5cm}
\end{center}
\caption{Binder cumulant data for the Gaussian interaction ISG in dimension 4. Sizes $L = 4,6,10$.}
\label{fig:3a}
\end{figure}

\begin{figure}
\begin{center}
\epsfig{file=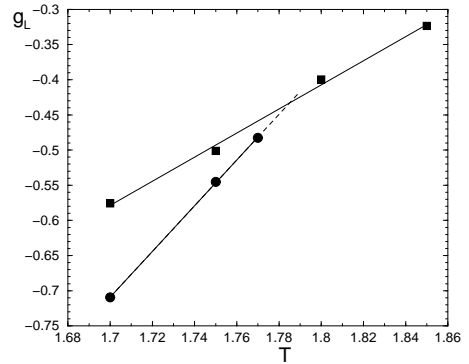,height=5cm}
\end{center}
\caption{$\eta(T)$ data for the Gaussian interaction ISG in dimension 4. Squares correspond to the SG susceptibility, cicles to the $(h,x)$ method.
The $\eta$ value at the intersection is clearly different from that of the 
binomial interaction distribution.}
\label{fig:3b}
\end{figure}

Figure~\ref{fig:3a} shows Binder cumulmant data for the 4d Gaussian case ; it can be seen
that the intersection point lies at $T=1.785 \pm 0.01$.  Figure~\ref{fig:3b} shows the
three scaling rule intersection.  Again the agreement is excellent, but the
value of $\eta$ at the intersection point $\eta = -0.44 \pm 0.02$, is
considerably higher than for the binomial case.  

\begin{figure}
\begin{center}
\epsfig{file=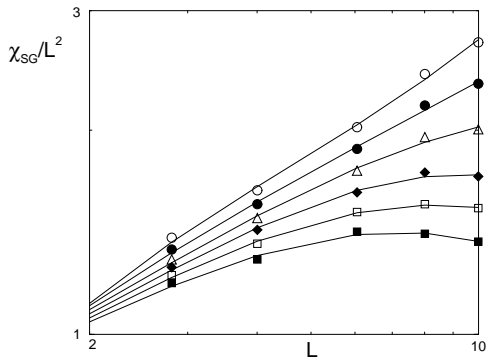,height=5cm}
\end{center}
\caption{The spin glass susceptibility $\chi_{SG}$ normalized by $L^2$ against 
$L$ on a log-log plot for the Gaussian interaction ISG in dimension 4. Temperatures from top to bottom are 1.75,1.77,1.79, 1.811.83,1.85.}
\label{fig:4}
\end{figure}

Figure~\ref{fig:4} shows a direct plot of$log(\chi_{SG}/L^2)$ against $log(L)$.  This has to 
be straight at $T_g$.  It
can be seen that $T_g$ must lie just below $T=1.79$, and $\eta$ can be estimated
from the corresponding slope.  We can note that there is excellent agreement
point by point between the data of \cite{ricci} and the present results wherever
comparisons can be made.  Using a scaling plot we find $\nu = 1.08 \pm 0.10$.

The 4d binomial exponent values given here are consistent with but more accurate
than those obtained in \cite{marinari}.  For the Gaussian system, the value of
$\eta= -0.35 \pm 0.05$ quoted in \cite{ricci} is low because of a marginally overestimated
$T_g$.

A careful analysis of the data on these two systems in dimension 4 but with
different interaction distributions shows conclusively that the exponents are
different, so universality does not hold.

In dimension 5, there is excellent agreement between series estimates
\cite{klein} and Monte Carlo estimates \cite{lwb2} for the binomial system.

\begin{figure}
\begin{center}
\epsfig{file=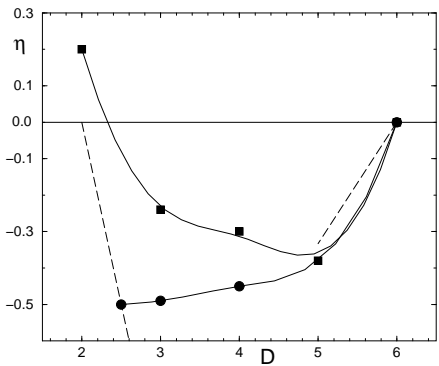,height=5cm}
\end{center}
\caption{The critical exponent $\eta$ as a function of dimension for ISGs 
with binomial (squares) and Gaussian (circles) interactions. The upper straight line represents the 
leading term in the $\epsilon$ expansion \protect\cite{green} starting from the upper 
critical dimension 6. The lower straight line is equation (10). The points in dimensions 3,4 and 5 are from the 
"three scaling rules" method. The dimension 2 point is from \protect\cite{BY}. 
The lower critical dimension point for the Gaussian interactions is explained 
in the text. }
\label{fig:5}
\end{figure}

In figure~\ref{fig:5} we display results for $\eta$ as a function of dimension for $\pm J$
and Gaussian interactions, including the point at $d_l$ for the Gaussian case
and the point at $d=2$ for the $\pm J$ case.  The point at $d=6$ corresponds to
the upper critical dimension value $\eta= 0$.  The straight line starting from 
this point
is the leading term in the $\epsilon$ expansion :  $\eta(d) = -(6-d)/3 +...$
\cite{green}.  It is clear that the data demonstrate that the exponents for the
two systems, binomial and Gaussian, follow regular curves as a function of
dimension.  Independent results are consistent with each other.  However the
values at each dimension are different for the two sets of interactions so
universality is clearly violated.  Results for other sets of interactions
\cite{lwb,lwb2} confirm this variation.

Critical exponents have also been measured in the Ising phase diagram as a
function of the ratio $p$ of the number of ferromagnetic to antiferromagnetic
bonds.  In the ferromagnetic region to the right of the Nishimori line, the
static exponents do not appear to vary with $p$, but the dynamic exponent $z$
does change continuously and very significantly \cite{ito}.

We can note that the Migdal-Kadanoff renormalization approach (which should be
exact for a heirachical lattice) has been used to measure effective ordering
temperatures and exponents for four different ISG interaction distributions.
\cite{nog}.  For dimension 3, a diamond hierachical lattice, and a
renormalization factor $b=2$, the ordering temperatures are in excellent
agreement with Monte Carlo estimates on cubic lattices.  However the
Migdal-Kadanoff calculations lead to a universal saddle point critical
temperature and exponent values.  This result seems to be closely related to the
hypotheses intrinsic to the Migdal Kadanoff method.

\section{Spin Glass Experiments}

Estimates of exponents have been made by many experimental groups, in general
using slightly different protocols and on different materials.  The non-linear
susceptibility is defined as \begin{equation} \chi_{nl}= \chi_0 - M/H
\end{equation} where $\chi_0$ is the zero field magnetic susceptibility.  At
$T_g$ $\chi_{nl}$ should behave as $H^{2/\delta}$ and above $T_g$
\begin{equation} 
M=\chi_0H +\chi_2 H^3 + \chi_4 H^5 +... 
 \end{equation} 
 with
$\chi_2$ diverging as $t^{-\gamma}$, where $t$ is the reduced temperature.  The
Suzuki equation of state is \cite{suz2} \begin{equation}
\chi_{nl}=t^{\beta}g(H^2/t^{\beta + \gamma}) \end{equation} Both d.c.  and a.c.
magnetization techniques have been used .  For instance Monod and Bouchiat
\cite{monod} and Bouchiat\cite{bou} used d.c.  entirely, taking care to stay in
a field range where the non-linear magnetization remained less than 10\% of
the linear magnetization so as not to corrupt the data.  Svedlindh et al
\cite{sved1} first analysed low field and low frequency a.c.  measurements to
fit to $\gamma$ and $T_g$.  (The ordering temperature was in agreement with the
low d.c.  field cusp temperature).  They then used the equation of state with
d.c.  measurements up to moderate fields with $\beta$ as the only fit parameter.
In a second set of experiments \cite{sved2} they measured the field and
frequency dependence of the a.c.  susceptibility and obtained an estimate of the
dynamics exponent $z\nu$.  In a sophisticated experiment, L\'evy \cite{levy}
measured the Fourier transform spectrum of the magnetization response to a 0.1
Hz field, picking up a series of non-linear susceptibilities from the different
harmonics.  He could deduce accurate values of static and dynamic exponents.

The experiments have to be performed with care, perticular attention being paid
to the proper identification of the critical temperature and to the necessity to
remain in a suitably low field range throughout.  Recent measurements
\cite{doro} in which the exponent $\delta$ was measured in a range of different
materials using one single protocol gave excellent confirmation of earlier data
by other teams (except for the case of {\bf Au}Fe where an early experiment had
given values of exponents out of line with all other results).  These results
validate the earlier measurements and show that the considerable spread of
values of exponents reported for different materials, is not due to
artefacts in the different measuring procedures.  As in the numerical data it is
evident that the expected universality of exponents breaks down.

The finite temperature $T_g$ values for real material spin glasses have been
somewhat of a mystery for some years.  These systems are Heisenberg, and
reliable numerical work has demonstrated that Heisenberg spin glasses in
dimension 3 should have zero temperature ordering\cite{olive,kawa}.
Kawamura\cite{kawa} has made the interesting suggestion that the ordering
process in real life Heisenberg materials is basically a chiral spin glass
ordering.  This ordering would not be visible directly to magnetization
experiments if there were no anisotropy.  However in all real systems random
anisotropy (of the DM type \cite{fert}) is always present, and by coupling the
chiral degrees of order with the magnetism an anisotropy, however weak, reveals
the chiral order.

The critical exponents for pure Heisenberg chiral ordering in dimension 3 have
been estimated numerically \cite{kawa}.  The best values are around $\nu = 1.25$
and $\eta = +0.7$.  It can be noted that $\nu$ is similar to the Monte Carlo
Ising values which we have presented, while $\eta$ is strongly positive rather
than negative as seen in the numerical Ising work.  A plausible hypothesis is
one in which the exponents change progressively from chiral-like for weak
anisotropy to Ising-like for strong anisotropy.  On this scenario, the value of
$\nu$ should remain relatively stable for all the materials, while the value of
$\eta$ should vary progressively, becoming gradually more negative as anisotropy
increases.

So far the qualitative trend for systems where both anisotropy and exponents
have been measured is in excellent agreement with this picture.  All the
experimental exponent values fall within the chiral limit at one end and the
Ising limit at the other end.  For the three alloy systems {\bf Ag}Mn, {\bf
Cu}Mn and {\bf Au}Fe the anisotropies are weak, moderate and strong
respectively.  The $\nu$ values are similar, near $1.3$, while the $\eta$ values
are about $0.4, 0.1$ and $-0.1$\cite{doro}.  The trend of exponent values is
clearly in the sense predicted by the scenario.

\section{Conclusion}

The main lesson which can be drawn from this overview of numerical and
experimental exponent data in spin glasses is that transitions in these glassy
systems are quite different from those in regular systems with standard second
order transitions.  The values of the exponents are far from those in regular
systems and the breakdown of Universality is manifest in carefully analysed
data.

The statistical physics community has been very loth to accept evidence
against universality because the renormalization scenario appears to give such
an appealingly general picture of behaviour at transitions.  However the fact
that it has proved extremely difficult on the field theory level to produce
predictions for spin glasses is a strong indication that unexpected behaviour
cannot be excluded {\it a priori}.

What possible mechanism could lead to this breakdown ?  It has been found
analytically that Ising spin systems with ferromagnetic interactions on
hierachical lattices show no universality \cite{hu}.  For the spherical model on
graphs of non-integer dimension, the exponents vary continuously with the
spectral dimension of the graph \cite{cassi}.  We can speculate that in spin
glasses the effective dimension of the system at criticality could depend on the
form of the interaction distribution.

It would be unfortunate if this phenomenon was left unexplored because of
preconceptions as to the physical laws which should hold for complex systems.
If it could be accepted that critical behaviour is much richer in glassy systems
than at conventional second order transitions without frustration, an important
new field of investigation should open out.  What control parameters affect the
exponents and why ?  What are the implications for the physics of the glass
transition in the most general sense ?










\end{document}